\documentclass[a4paper,11pt]{article}
\pdfoutput=1
\usepackage[utf8x]{inputenc}

\usepackage{color,graphicx,dsfont,slashed,subfigure,multirow,enumitem,jheppub}
\hypersetup{bookmarks=true,unicode=true,pdftoolbar=true,pdfmenubar=true,
	pdffitwindow=false,pdfstartview={FitH},pdftitle={Combine and Conquer: Event Reconstruction with Bayesian Ensemble Neural Networks },
	pdfauthor={J.~Araz,M.~Spannowsky}, pdfsubject={Ensemble Neural Networks},
	pdfcreator={J.~Araz,M.~Spannowsky},pdfproducer={texstudio}, pdfkeywords={deep learning},
	pdfnewwindow=true,colorlinks=true,
	linkcolor=blue,citecolor=magenta,filecolor=magenta,urlcolor=cyan}
\definecolor{lightgrey}{gray}{0.9}

\def\btabu#1\etabu{\begin{tabular}{p{125mm}}#1\end{tabular}}
\def\btab#1\etab{\begin{tabular}{p{50mm}p{70mm}}#1\end{tabular}}
\def\btabnn#1\etabnn{\begin{tabular}{p{45mm}p{75mm}}#1\end{tabular}}
\def\btabx#1\etabx{\begin{tabular}{p{65mm}p{55mm}}#1\end{tabular}}
\def\btaby#1\etaby{\begin{tabular}{p{40mm}p{80mm}}#1\end{tabular}}
\def\btabyy#1\etabyy{\begin{tabular}{p{20mm}p{100mm}}#1\end{tabular}}
\def\btabzz#1\etabzz{\begin{tabular}{p{35mm}p{85mm}}#1\end{tabular}}
\def\btabyyy#1\etabyyy{\begin{tabular}{p{10mm}p{110mm}}#1\end{tabular}}
\def\btabyyyy#1\etabyyyy{\begin{tabular}{p{2mm}p{118mm}}#1\end{tabular}}
\def\btabwide#1\etabwide{\begin{tabular}{p{82mm}p{38mm}}#1\end{tabular}}
\def\bcen{\begin{center}}
	\def\ecen{\end{center}}
\def\bgfb#1\egfb{\bcen\fcolorbox{black}{lightgrey}{\parbox{130mm}{\btabu#1\etabu}}\ecen}
\def\bgfbn#1\egfbn{\bcen\fcolorbox{black}{lightgrey}{\parbox{130mm}{\btab#1\etab}}\ecen}
\def\bgfbnn#1\egfbnn{\bcen\fcolorbox{black}{lightgrey}{\parbox{130mm}{\btabnn#1\etabnn}}\ecen}

\def\bgfbx#1\egfbx{\bcen\fcolorbox{black}{lightgrey}{\parbox{130mm}{\btabx#1\etabx}}\ecen}
\def\bgfbyy#1\egfbyy{\bcen\fcolorbox{black}{lightgrey}{\parbox{130mm}{\btabyy#1\etabyy}}\ecen}
\def\bgfbyyyy#1\egfbyyyy{\bcen\fcolorbox{black}{lightgrey}{\parbox{130mm}{\btabyyyy#1\etabyyyy}}\ecen}
\def\bgfbzz#1\egfbzz{\bcen\fcolorbox{black}{lightgrey}{\parbox{130mm}{\btabzz#1\etabzz}}\ecen}
\def\bgfbyyy#1\egfbyyy{\bcen\fcolorbox{black}{lightgrey}{\parbox{130mm}{\btabyyy#1\etabyyy}}\ecen}

\def\bgfbalign#1\egfbalign{\bcen\fcolorbox{black}{lightgrey}{\parbox{130mm}{\btaby#1\etaby}}\ecen}
\def\bgfbwide#1\egfbwide{\bcen\fcolorbox{black}{lightgrey}{\parbox{130mm}{\btabwide#1\etabwide}}\ecen}

\newcommand{\be}{\begin{equation}}
\newcommand{\ee}{\end{equation}}
\def\bsp#1\esp{\begin{split}#1\end{split}}
\def\bpm{\begin{pmatrix}} 
	\def\epm{\end{pmatrix}} 
\renewcommand{\figureautorefname}{Fig.}

\def\sectionautorefname~#1\null{Sec.~#1\null}
\def\subsectionautorefname~#1\null{Sec.~#1\null}
\def\figureautorefname~#1\null{Fig.~#1\null}
\def\tableautorefname~#1\null{Table~#1\null}
\def\equationautorefname~#1\null{Eq.~(#1)\null}

\preprint{IPPP/20/74}

\begin{document}
	
	\date{\today}
	
	\title{Combine and Conquer: Event Reconstruction with Bayesian Ensemble Neural Networks}
	
	\author[a]{Jack~Y.~Araz}
	\author[a]{and Michael~Spannowsky}
	\affiliation[a]{Institute for Particle Physics Phenomenology,\\Durham University, South Road, Durham, DH1 3LE,}
%
	\emailAdd{jack.araz@durham.ac.uk}
	\emailAdd{michael.spannowsky@durham.ac.uk}

	\vspace{10pt}
	\abstract{
		Ensemble learning is a technique where multiple component learners are combined through a protocol. We propose an Ensemble Neural Network (ENN) that uses the combined latent-feature space of multiple neural network classifiers to improve the representation of the network hypothesis. We apply this approach to construct an ENN from Convolutional and Recurrent Neural Networks to discriminate top-quark jets from QCD jets. Such ENN provides the flexibility to improve the classification beyond simple prediction combining methods by linking different sources of error correlations, hence improving the representation between data and hypothesis. In combination with Bayesian techniques, we show that it can reduce epistemic uncertainties and the entropy of the hypothesis by simultaneously exploiting various kinematic correlations of the system, which also makes the network less susceptible to a limitation in training sample size.
	}

	
	\maketitle

	\section{Introduction}\label{sec:intro}
	
	Deep Learning (DL) has gained tremendous momentum on the verge of the latest developments in data analysis. Whilst boosted decision trees (BDT) have been used in the context of High-Energy Physics for over 30 years, wide usage of Deep Neural Networks (DNNs) only surged very recently. Since then, especially in applications to LHC physics where a large amount of data with the need for its fast and automated analysis is gathered, there has been a profound improvement in the understanding of Neural Networks (NNs). The analysis of the internal structure of jets, highly complex collimated sprays of radiation \cite{Marzani:2019hun}, is a popular arena where reconstruction techniques evolved from sophisticated multi-variate approaches, e.g. \textsc{HEPTopTagger}~\cite{Plehn:2009rk, Plehn:2010st, PhysRevD.85.034029}, over theory-guided matrix-element methods \cite{Soper:2012pb,Soper:2011cr,Soper:2014rya,Prestel:2019neg} to data-driven NN techniques \cite{Brehmer:2018kdj,Brehmer:2019xox,Louppe:2016ylz,Khosa:2020qrz}. In particular top tagging has been the prime example to benchmark the performance of various NN classifiers~\cite{Almeida:2015jua, Kasieczka:2017nvn, Butter:2017cot, Pearkes:2017hku, Egan:2017ojy, Macaluso:2018tck, Choi:2018dag, Moore:2018lsr,Blance:2019ibf}. Similar tagging algorithms have been used for Higgs~\cite{Lim:2018toa, Lin:2018cin} and W-boson~\cite{Baldi:2016fql, Louppe:2017ipp} tagging and quark-gluon discrimination~\cite{Gallicchio:2012ez, Komiske:2016rsd, Cheng:2017rdo, Komiske:2018cqr, Bright-Thonney:2018mxq}\footnote{For a review of these methodologies and more see refs.~\cite{Larkoski:2017jix, Kasieczka:2017nvn}, and other examples~\cite{deOliveira:2015xxd, Kitouni:2020xgb, Ju:2020tbo, Butter:2020qhk, Farrell:2019fsm, Lin:2019htn, Datta:2019ndh,DAgnolo:2019vbw,DAgnolo:2018cun,Nachman:2021yvi, Faucett:2020vbu, Khosa:2019qgp, Khosa:2019kxd}.}. Thus, it became apparent that there is a wide range of use-cases for NNs in collider phenomenology, where particle tagging is just one of many applications.


	A standard supervised learning algorithm produces a fitting function that aims to find an optimal contour of the decision boundary between competing hypotheses\footnote{Here the word ``fitting" is used to simplify the text. However, Deep Learning is not merely a fitting algorithm; it looks for a higher dimensional irreducible representation that the feature-space lives in.}. The given algorithm takes a labelled feature-tensor and attempts to find the global minimum of a given objective function, the so-called loss function, resulting in the prediction of the algorithm. This is achieved by convoluting the input feature vector with non-linear functions, so-called activation functions, and updating the weights of the initial hypothesis through the backpropagation algorithm. Whilst such an approach offers increased flexibility, in general, it can suffer from three major predicaments~\cite{10.1007/3-540-45014-9_1}. First, the problem of statistics denotes the lack of training examples within a particular domain, which can cause the learning algorithm to get stuck in various minima with comparable accuracies in each training. 
	The second problem is computational. As mentioned before, a learning algorithm often employs a stochastic search algorithm, e.g. gradient descent. Assuming the provision of sufficient data, the feature-space can be highly complex, creating a very non-trivial loss-hypersurface for which the algorithm is tasked to find the global minimum~\cite{58871, BLUM1992117}. Finally, the third problem is representational. As the nature of a ``fitting" algorithm, it is not always possible to find a linear or non-linear representation of the actual function. Hence, it might be necessary to expand the representation space or employ various possible hypotheses to find a closer approximation of the actual function. Although the representation problem is directly linked to the previously mentioned issues, even with sufficient statistics and advanced algorithms, an optimization algorithm may not proceed after finding a hypothesis that can adequately explain the data~\cite{HORNIK1990551}.

	The three most popular architectures for classification tasks in particle physics are currently Deep Neural Networks (DNN), Convolutional Neural Networks (CNN) and Recurrent Neural Networks (RNN). Each of these networks is designed to exploit different features and correlations of the input data. For instance, 
	CNNs are special-purpose networks that are widely used for image recognition~\cite{Kasieczka:2017nvn, Macaluso:2018tck}. This method sweeps through the image by dividing it into subvolumes. Each subvolume has been transferred to the next layer by passing through an activation function, allowing the network to filter the image's distinguishable features. RNNs are a different kind of specialised networks that keep track of the ordering of the feature vector and thus maintains a sense of ``memory" by connecting each node in a graph via an ephemeral sequence. Long-short term memory (LSTM) networks have been employed to classify QCD events with high accuracy~\cite{Louppe:2017ipp, Egan:2017ojy,Englert:2020ntw}. While each of these techniques can be powerful by itself, it is not clear whether they exploit the full amount of information contained in the feature vectors to perform an optimal classification between competing hypotheses. Thus, combining multiple networks into an Ensemble Neural Network (ENN) might allow to improve on their individual classification performances.
	
	Ensemble learning is a paradigm which employs multiple neural networks to solve a problem. The main idea behind ensembling is to increase the generalisation of the data by harvesting many hypotheses trying to solve the same problem. Each of the networks mentioned above is specialised to learn a particular feature of the given data to achieve the same or similar generalisation. An ensemble of these networks can access all the information presented in each component network and optimise it according to more generic information through data~\cite{ZHOU2002239, 10.5555/2998687.2998716, doi:10.1142/9789812795885_0025, xie2013horizontal, Rokach:2010aa, 10.1007/3-540-45014-9_2}. 

	While some techniques to combine classifiers have been used in the context of collider phenomenology before\footnote{Ref.~\cite{Conrad:2006ip} shows that combining predictions of BDTs with specific rules can improve the discrimination of BSM models from the SM. Ref.~\cite{Baldi:2014pta} shows that injecting randomness to a hypothesis and combining its results can boost the accuracy of the classification. Refs.~\cite{Alves:2016htj,Alves:2019ppy} uses stack combining method for Higgs tagging at LHC and ref.~\cite{Kasieczka:2019dbj} combines the predictions of multiple different learners.}, to our knowledge for the first time, we will present a parallel combining method to go beyond simple prediction combinations.  As shown in previous studies~\cite{58871,10.1007/3-540-45014-9_1,10.5555/2998687.2998716,548872}, combining predictions of various networks can significantly improve the overall performance for classification or regression. However, if networks are only combined at the prediction level, they are each separately trained for a specific property of the data. Parallel combined ensembles allow the network to train on a combined higher dimensional latent-space to optimize the entire network accordingly. Hence, having access to all component networks allows improvement upon the representation of the problem. We will show that such an approach allows flexibility to improve background rejection beyond simple prediction combinations. Furthermore, we will show that it will drastically improve the network's error correlations beyond the component and prediction-based-combined networks.
	
	With continuously improving performance indicators for NNs, e.g. measured through receiver operating characteristic (ROC) curves, it becomes increasingly important to obtain an understanding of how this is achieved and how reliable the performance is evaluated~\cite{Bollweg:2019skg, 2018DPS....5050501M, DBLP:journals/corr/abs-1811-09385,Nachman:2019dol,Nachman:2020fff,Englert:2018cfo}. Bayesian neural networks allow to estimate intrinsic uncertainties of NN by treating their weights as distributions instead of a single trainable variable~\cite{gal2016dropout,kendall2017uncertainties}. Hence the network output is a distribution rather than a fix value. To estimate the uncertainties of a network, multiple measurements of the same test data are combined to calculate the mean prediction alongside its standard deviation. We will employ Bayesian techniques to show that parallel combining methods, i.e. as implemented in ENNs, can reduce the standard deviation of the predictions and epistemic uncertainties without requiring more data.
	
	
	In \autoref{sec:enns} we provide a discussion of Ensemble Neural Networks and review their applications and benefits in improving the classification performance.  In \autoref{sec:preprocess} we describe the procedure we employed to preprocess the input data before the training and in \autoref{sec:nn_achi} we present our results. Finally, in \autoref{sec:bnn} we compare uncertainties between component networks and their ensemble, and we offer a summary and conclusions in \autoref{sec:conclusion}.

	\section{Ensemble Neural Networks}\label{sec:enns}
	Ensemble Neural Networks (ENNs) are protocols that aim to increase the generalizability of a hypothesis by combining multiple component networks. It has been shown that ENNs can provide the necessary resolutions or approximations that that all three potential pitfalls for NNs mentioned in \autoref{sec:intro} require~\cite{ZHOU2002239,10.5555/2998687.2998716,doi:10.1142/9789812795885_0025,xie2013horizontal, Rokach:2010aa}. Depending on the problem at hand, ensembling methods can be pooled under three paradigms~\cite{10.1007/3-540-45014-9_2}:  (i) parallel combining, (ii) stacked combining and (iii) combining weak classifiers.
	
\begin{figure}[!h]
		\centering
		\includegraphics[scale=1.]{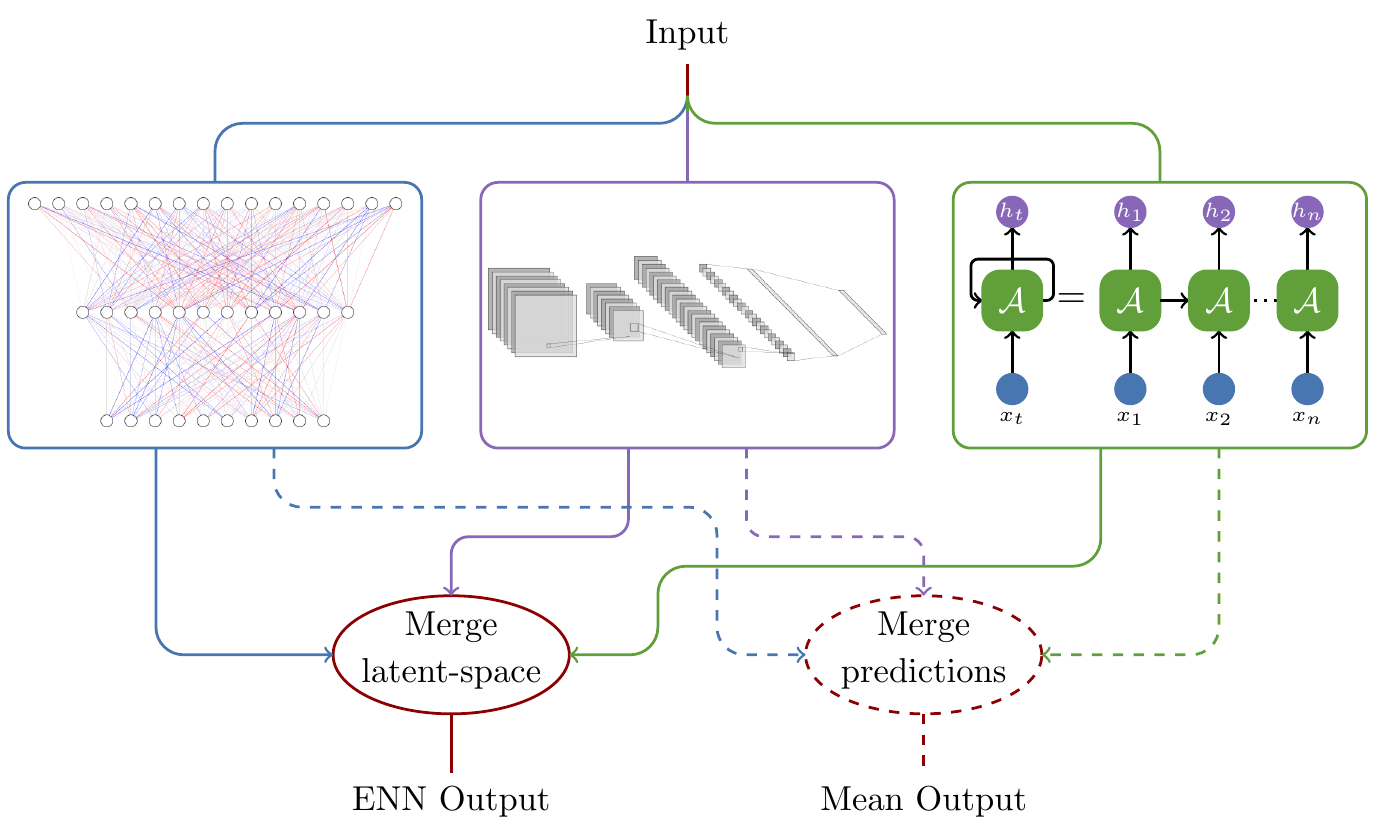}
		\caption{\it A schematic representation of ensemble neural networks where blue box represents a NN with dense layers, purple represents convolutional 	neural network and green represents a recurrent NN with inputs $x_i$ and output values $h_i$ for an operator $ \mathcal{A} $. Solid line at the bottom guides towards latent-space concatenation which leads to ensemble prediction. Dashed lines represent the same for mean prediction of each network.} \label{fig:enn}
	\end{figure}

	Combining classifiers spanning feature-spaces that contains different physical domains, can provide an expanded representation of the hypothesis space, see \autoref{fig:enn}. Such methods are studied under so-called ``parallel combining" method. Another technique, called ``stacked combining", employs different classifiers to be trained on the same feature-tensor. Such techniques can provide simple solutions to the computational problem where multiple non-correlated hypotheses can approximate the underlying function more efficiently. The final and most widely studied method is ``combining weak classifiers" where, as the name suggests, weak but successful classifiers' predictions are assembled to create a NN that reaches accuracies beyond its constituents~\cite{10.1007/3-540-45014-9_1}. Here successful means that the hypothesis has greater accuracy than random selection.  Although existing methods under the paradigms $(ii)$ and $(iii)$ can successfully optimize over statistical and computational shortcomings of the NNs~\cite{10.5555/2986766.2986883,Cherkauer96humanexpert-level,doi:10.1080/095400996116839, Breiman:1996aa, gams1989new, NIPS1995_46072631,Freund:1997xna, Freund96experimentswith}, they can not expand the representation of the hypothesis without acquiring an extended domain of the data. Hence one needs a dedicated approach to address the representation problem to learn over different types of correlations within distinct symmetries of the data.

	While ENNs are known to improve on the statistics and computational problems~\cite{10.1007/3-540-45014-9_2}, see \autoref{sec:intro}, its benefits for the representation problem, which is in most collider phenomenological applications often is prevalent, is underrated.  We propose the use of ENNs for the event reconstruction at high-energy collider experiments under the paradigm of parallel combining. We will further show that this approach improves on the representation problem.




	For this purpose, we will use two high-level classifiers, a CNN and a RNN which are often used for image recognition and text recognition respectively. Both of these models are generalising a specific property of a jet, {\it i.e.} the spatial position of the substructure of a jet and the sequential order of a cluster history respectively. Naively, one could take the mean prediction of both classifiers, which will lead to a generalisation of the problem in the higher-dimensional hypothesis space. Although this can improve the performance, both component networks are optimised for their own feature space. In this study, we show that instead of combining the component networks' predictions, optimising the network over the combined latent-feature space can lead to a more substantial and stable performance improvement for the problem at hand. 

	Thus, we propose to initialise multiple high-level classifiers separately. For the example of \autoref{sec:tagging}, these are chosen to be CNN and RNN classifiers. Each the CNN and RNN provide a vector in the latent-feature space corresponding to the flattened image for the CNN and the higher dimensional representation sequence for the RNN. Concatenating these vectors will lead to a larger latent-space, including information from both image-type and sequence-type data. 
	Training with this higher dimensional feature space with extra handles for the NN architecture, such as more layers or nodes to generalise this latent-feature space, can lead to two significant improvements. Firstly, each component network's weights will be optimised with respect to the combined hypothesis space hence will have access to more features of the base theory. Secondly, the ability to access a larger latent-feature space will make it possible to increase the complexity of the model for a larger hypothesis-space. 

	\autoref{fig:enn} shows a schematic representation of this approach where one source of input is divided into multiple branches to be analysed within different architectures. Depending on the nature of the problem, one can employ multiple network architectures such as fully connected networks (blue), CNNs (purple), RNNs (green) or even more complex structures which, for the sake of simplicity, are not shown explicitly. The merging stage represents the concatenation process where instead of the prediction of each model, one can combine the latent-space of each network after its individual $i^{th}$ layer and continue training on this new feature space. Hence, the network's output will be the prediction optimised over each distinct feature of the problem.
	
	Whilst the network architectures discussed often unveil a strong performance improvement over conventional cut-based reconstruction strategies; one wonders if combining any NN will increase accuracy. To answer this question one needs to investigate the bias-variance-covariance decomposition. The prediction of an ensemble estimator, constructed by averaging the prediction of each component estimators, assuming that they are independent from each other, can be cast as 
	\begin{eqnarray}
	f_{ens}(\mathbf{x}) = \frac{1}{N}\sum_{i}^{N}f_i(\mathbf{x})\ ,
	\end{eqnarray}
	where N is the number of component estimators, $f_i(\mathbf{x})$ is the prediction of the $i^{\rm th}$ estimator and $\mathbf{x}$ is the feature-tensor. For such an object, the generalization error is given by~\cite{548872,DBLP:journals/jmlr/BrownWT05}
	\begin{eqnarray}
	\mathrm{Err}(f_{ens}) = \mathrm{Err}\left\{ \frac{1}{N}\overline{\mathrm{Var}}(\mathbf{x}) + \left(1-\frac{1}{N}\right)\overline{\mathrm{Cov}}(\mathbf{x}) + \overline{\mathrm{Bias}}(\mathbf{x})^2 \right\}\ ,
	\end{eqnarray}
	where the three terms correspond to variance, covariance and the squared bias of the feature-tensor respectively. Although such construction assumes a very simplistic case, it shows that the generalization error of the average prediction of multiple hypotheses is also affected by the covariance. This shows that if the component hypotheses are negatively correlated with each other the average prediction will decrease the generalization error further. However, as the average bias will remain the same, the generalization error can only be reduced to the bias term. Thus an ENN can improve the generalization error if and only if the given component estimators' errors are not completely correlated~\cite{ZHOU2002239, 10.5555/645529.657784}.
	
	\section{Top Tagging through Ensemble Learning}\label{sec:tagging}

	Using CNNs, the pixelated energy deposits in the calorimeters of multi-purpose high-energy experiments have been repeatedly shown to provide a strong discriminatory power between the radiation profile of top quarks versus QCD jets. In the $\eta-\phi$ plane, each pixel corresponds to one or more particles, and so-called colour or luminosity of a pixel can be measured by a combined intrinsic property of these particles such as energy or transverse momentum. This will allow the CNN to learn translationally invariant features of the top and jet system. RNNs instead maintain a sense of timing and memory in a given sequence used as input features. Due to the nature of the clustering algorithm, each jet has an embedded tree structure, where subjets are recombined with respect to a particular rule. Thus, CNNs and RNNs exploit different features of top and QCD jets to discriminate them from each other. We use the complementarity of both methods to combine them in an ENN that has an improved performance over both approaches individually. An implementation of the code we use for preprocessing and network training is provided at \href{https://gitlab.com/jackaraz/EnsembleNN}{this link}\footnote{\href{https://gitlab.com/jackaraz/EnsembleNN}{https://gitlab.com/jackaraz/EnsembleNN}}.
	
	\subsection{Dataset \& Preprocessing}\label{sec:preprocess}
	
	As a case study, we will investigate the top tagging capabilities of NNs by employing a CNN and a RNN.  To achieve this, we used the dataset provided in \cite{Kasieczka:2019dbj, kasieczka_gregor_2019_2603256}, which consists of 14 TeV top signal and mixed quark-gluon background jets generated and showered by \textsc{Pythia}~8~\cite{Sjostrand:2014zea}. The detector simulation for showered events is obtained through the \textsc{Delphes}~3 package~\cite{deFavereau:2013fsa} using the default ATLAS detector card. The fat jets are reconstructed using \texttt{anti-kT} algorithm~\cite{Cacciari:2008gp} as defiend in \textsc{FastJet}~\cite{Cacciari:2011ma}, using radius variable $ R=0.8 $. The fat-jet transverse momentum has been limited to $ [550,650] $~GeV range in order to be able to benchmark the NN architectures precisely depending on the nature of jet substructure within a specific $p_T$-range. The resulting fat jets are further limited to be within $ |\eta_j|<2 $. Finally, the fat jets in the top signal sample have been matched with truth level tops requiring $ \Delta R(j,t_{truth})<0.8 $. This dataset consists of 1.2 million training, 400,000 validation and test events respectively.  This dataset has been divided into two subsets within our framework, one for CNN type training and one for RNN type training. For both of the datasets provided PFlow-objects are clustered into a fat-jet as described above. 
	
\begin{figure}[!h]
		\includegraphics[scale=0.45]{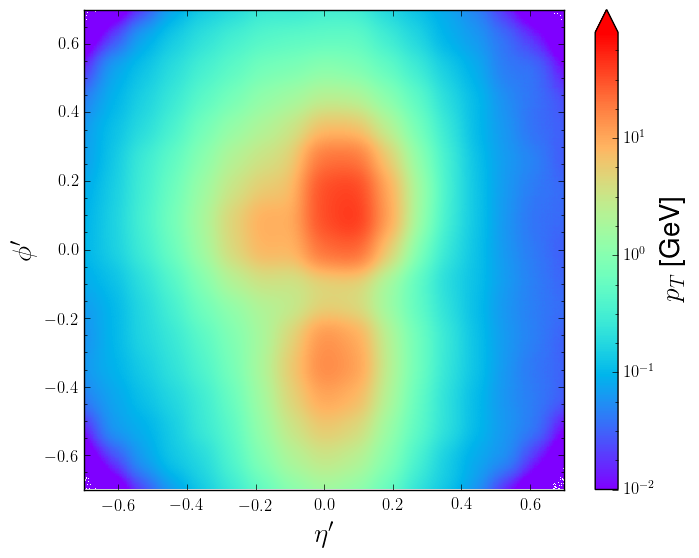}
		\includegraphics[scale=0.45]{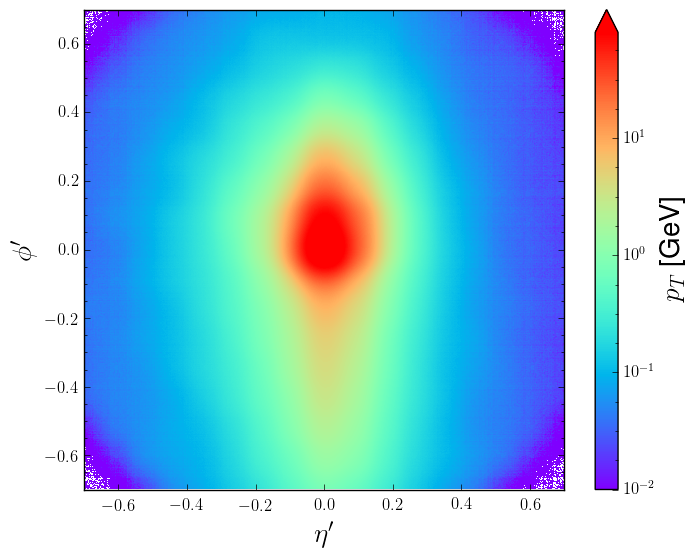}
		\caption{\it Left panel shows averaged top signal image on modified $ \eta-\phi $ plane and the left panel shows the same for dijet sample. Colour represents the combined transverse momentum of the constituents within a pixel. Each image includes 10,000 events.} \label{fig:preprocess_image}
	\end{figure}

	The CNN dataset has been prepared with leading \texttt{anti-kT} fat jet constituents which are ordered by their corresponding transverse momentum. Each jet in the event has been centred with respect to total $ p_T $ weighted centroid where the jet vector has been centred at $ (\eta,\phi) = (0,0)$. Furthermore, the coordinate system has been rotated such that the principal axis is at the direction of positive pseudo-rapidity for all constituents. These modified constituents are fitted into pixels on $ \eta-\phi $ plane, divided into $ 37\times37 $ pixels between $ (\eta,\phi) = ([-1.5,1.5], [-1.5,1.5])$ where the pixel value has been set as total $ p_T $ within that pixel. To get the leading constituent into the first quadrant, the vertical half of the image with higher total $ p_T $ flipped to the right, and similarly, the horizontal half of the image with higher $ p_T $ flipped to the top. \autoref{fig:preprocess_image} shows the averaged top signal (left) and dijet background (right) images for $ 10,000 $ events projected on modified $\eta-\phi$--plane. Colour represents the magnitude of the transverse momentum in the corresponding pixel. Note, this image has been zoomed-in to highlight the relevant portion of the image. Since the network requires the input data within $[0,1]$ range, each image has been normalized by 1 TeV before training.

	\begin{figure}[!h]
		\centering
		\includegraphics[scale=1.]{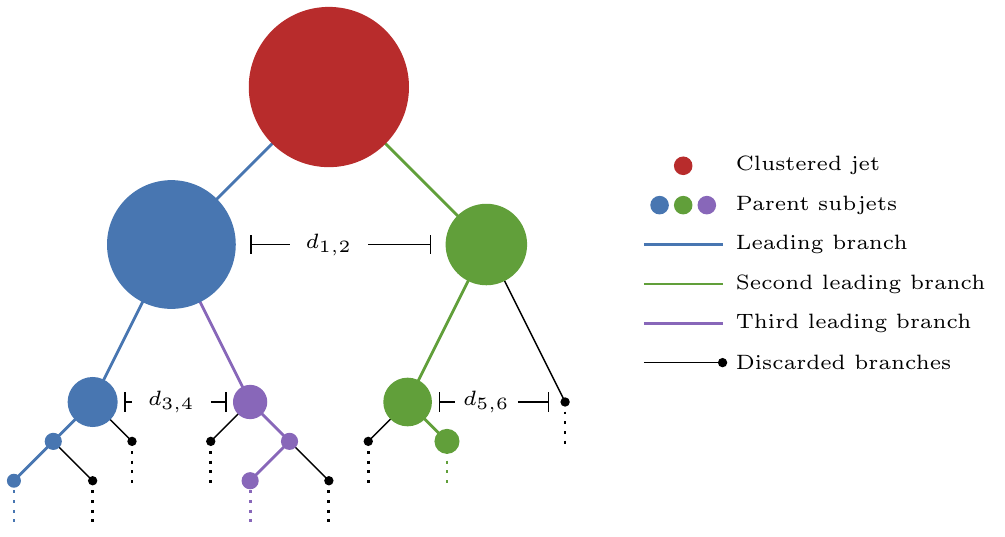}
		\caption{\it A schematic representation of the cluster history where blue  represents leading branch with respect to the relative magnitude of transverse momentum, green is the second leading branch and purple is the third leading branch. Black lines shows the discarded branches. Finally dark red represents the initial clustered jet. The size of the circles represents the relative magnitude of transverse momentum. } \label{fig:cluster_hist}
	\end{figure}
	\begin{figure}[!h]
		\centering
		\includegraphics[scale=.45]{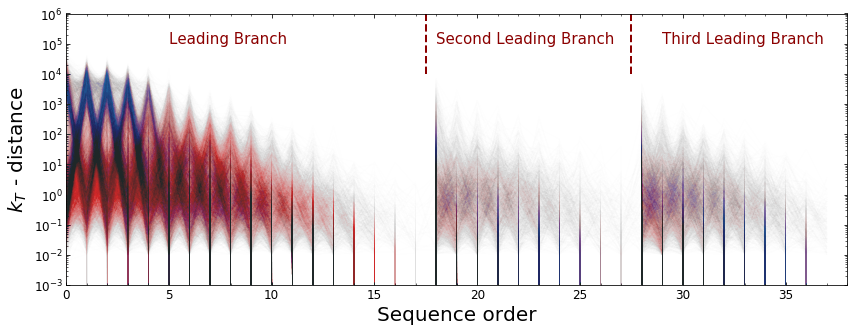}\\
		\includegraphics[scale=.45]{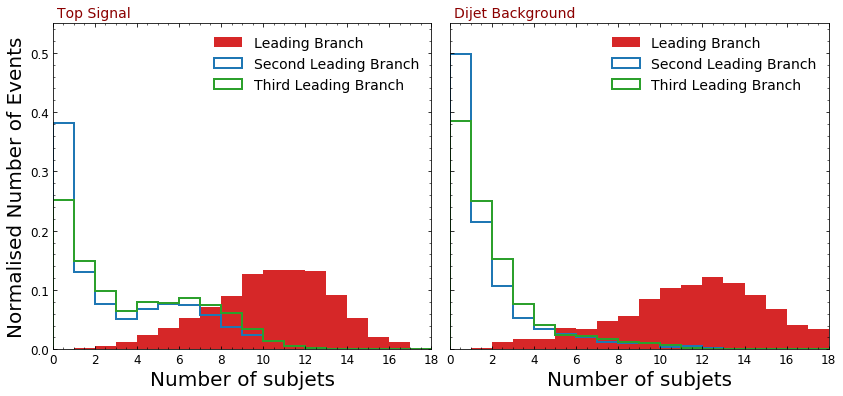}
		\caption{\it Top panel shows combined $ k_T $-distances in RNN sequence for 4000 events. Red represents the dijet events and blue represents top signal events. Dominated colours shows which event has high occurance in a particular sequence. Bottom two panel shows the number of subjets in each branch where left panel shows it for top signal and right panel shows for dijet background.} \label{fig:kTsequence}
	\end{figure}
	
	The RNN dataset has been constructed using leading \texttt{anti-kT} fat-jet where the constituents of the this jet are re-clustered with the  same radius parameter using the \texttt{Cambridge\-/Aachen} (\texttt{C/A}) clustering algorithm~\cite{Bentvelsen:1998ug}. In order to construct the training sequence, three leading branches have been extracted from the clustering history where their respective transverse momentum defined the branches. Initial two leading branches are constructed by the first two subjets in the clustering history where the subjet with larger $p_T$ has been chosen to be the leading branch. The third leading branch has been chosen within the parent subjets of the first leading subjet. The parent with the lowest $p_T$ is considered as the third leading branch. \autoref{fig:cluster_hist} shows a schematic representation of this selection where blue stands for the leading branch following the subjets with relatively higher momentum than the consecutive parent subjet. Green is the second leading branch and purple is the third leading branch following the same pattern as the leading branch. Black lines represent the discarded branches which have less $p_T$ compared to the corresponding parent subjet. Finally, red represents the \texttt{C/A}-jet. The sequence has been constructed using $k_T$-distances in the clustering history, defined as
	\begin{eqnarray}
		d_{i,j} = \min\left(p^2_{T,i},p^2_{T,j}\right)\frac{\Delta R^2}{R} \ .\nonumber
	\end{eqnarray}
	Here $i,j$ is the number of the parent subjets, $\Delta R$ is the relative angular distance between two subjets and R is the clustering radius given as 0.8. For each parent subjet in a branch, the $d_{i,j}$ value is stored with its chronological order. $d_{1,2}$ and $d_{3,4}$, see \autoref{fig:cluster_hist}, are included as part of the leading branch sequence. In order to compose the RNN sequence, we first used the mass of the \texttt{anti-kT}-jet and then the mass of  \texttt{C/A}-jet constructed using \textsc{Mass Drop Tagger}~\cite{Butterworth:2008iy}  ($ \mu = 0.8 $, $ y_{\rm cut} = 0.09 $).  Then we added the first 18, 10, 10 $k_T$-distances of the leading, second leading and third-leading branches, respectively. Branches with fewer subjets then padded with zeros. Upper panel of \autoref{fig:kTsequence} shows the $k_T$ sequence for 2000 top signal and 2000 dijet background events. Each event has been represented via high transparency; hence the vibrant colours show the high occurrences of the particular events where blue and red stands for top and dijet samples. The bottom two panels of \autoref{fig:kTsequence} show the number of subjets in each branch where the left and right panels show for top and dijet samples, respectively\footnote{It is important to note that we also test our sequence by constructing it out of jets clustered by \texttt{kT} and \texttt{anti-kT} algorithms; however, the discriminative power has been observed to be less than the sequence clustered by \texttt{C/A} algorithm.}. Before passing the input feature vectors to the network for training, the dataset has been standardized using \texttt{RobustScaler} within \textsc{Scikit-Learn} package~\cite{scikit-learn} using 100,000 mixed events from the training sample.

	\subsection{Network Architecture \& Training}\label{sec:nn_achi}

	In order to study the effects of ensembling multiple architectures, here we will first introduce two ``comparable" but independent architectures for the CNN and RNN-type of datasets presented in \autoref{sec:preprocess}.  Our NN architecture relies on \textsc{Keras} library~\cite{chollet2015keras} embedded in \textsc{TensorFlow} version 2.2~\cite{tensorflow2015-whitepaper}.

	The CNN dataset has been trained by a network receiving $ 37\times37 $-pixel input via a 2D convolutional layer with eight features and four stride pixels alongside with zero paddings. This layer's output is normalized within a batch normalization layer and passed on to a max-pooling layer with a pool size of $2\times2$, leaving a reduced $18\times18$ image with eight features. Finally, these images have been flattened and passed to a fully connected dense layer with sixteen nodes with a dropout probability of $25\%$. A rectified linear unit (\texttt{ReLu}) activation function has been used for each layer. A dense output layer has then followed the network with a single node and sigmoid activation for classification.

	Furthermore, the RNN dataset has been trained in a slightly more complex architecture starting with an LSTM layer, including 64 nodes. The activation and recurrent activation function for the LSTM layer have been chosen as hyperbolic tangent and sigmoid functions. It has been followed by three fully connected dense layer with 64, 64 and 32 nodes respectively and each dense layer followed by a dropout layer with $25\%$ probability. As before, the \texttt{ReLu} activation function has been used for each dense layer. The network output has been generated from a final dense layer with a single node and sigmoid activation function.

	Both networks are aimed to minimize a binary cross-entropy loss function via \texttt{Adam} optimizer~\cite{Kingma2014AdamAM} with a learning rate of $10^{-4}$. Networks are trained for 500 epochs, and the learning rate has been reduced half for every 20 epochs if there is no improvement on the validation dataset's loss value. If the network didn't improve the validation loss for 250 epochs, the training terminated automatically. 

	Since the goal of this study is to question if a more extensive representation can generalize the given problem much better than its component hypotheses, we employed two types of ensembling methods. As a reference case, we studied the mean of both CNN and RNN predictions. As mentioned in \autoref{sec:intro}, such ensembles have shown to go beyond the accuracies of their component networks. For the main case, we will study an extended architecture where CNN and RNN architectures are concatenated before their output layer; hence resulting in a latent-space of 48 features. To find an optimal generalization of this latent-feature space, they are further connected to a fully connected dense layer with 96 nodes, employing \texttt{ReLu} activation function and L2 kernel regularization with a penalty strength of 0.05. This dense layer has been padded with 25\% dropout layers before and after. Then connected to an output layer as before, activated via a sigmoid function.
	\begin{figure}
		\centering
		\includegraphics[scale=0.35]{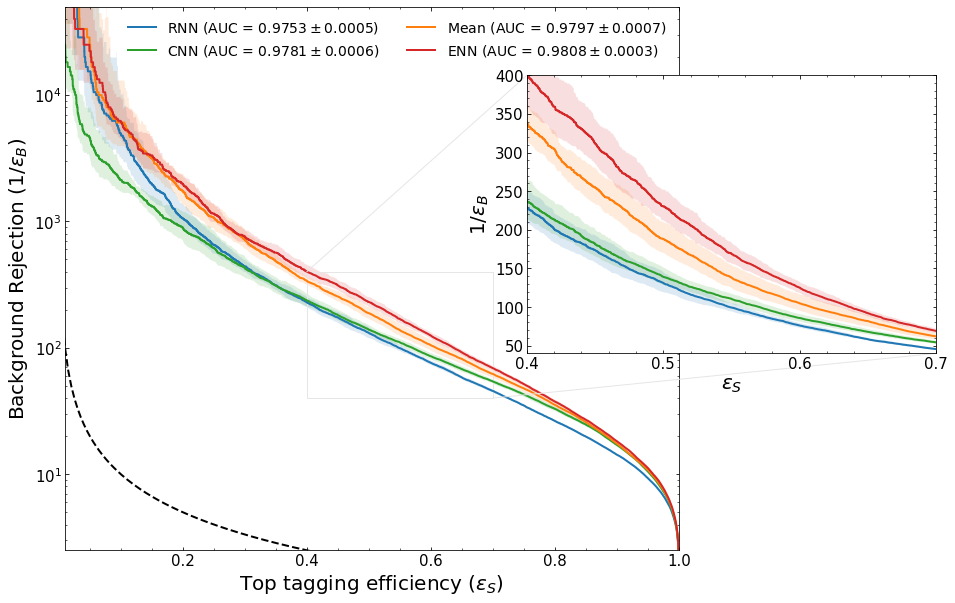}
		\caption{\it Receiver operating characteristic curve has been shown where CNN, RNN, the mean prediction of both and ENN architectures represented by green, blue, orange and red curves. The epistemic uncertainty has been represented by the transparent area around each curve for one standard deviation. Black curve represents the random choice. The inner plot zooms into the slice of $ \varepsilon_S\in [0.4,0.7] $.}\label{fig:roc}
	\end{figure}
	
	\begin{figure}
		\centering
		\includegraphics[scale=0.5]{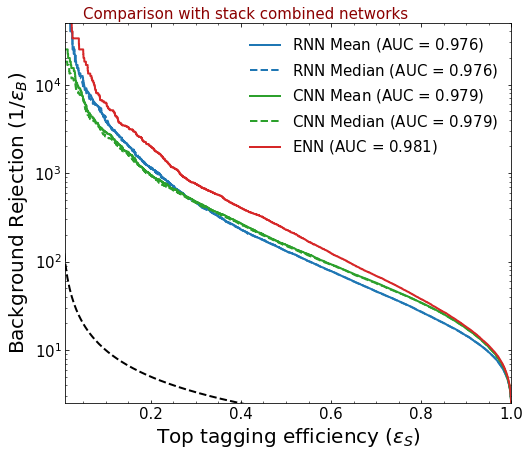}
		\caption{\it Comparison between stack and parallel combined networks have been shown. The receiver operating characteristic curve for CNN (green) and RNN (blue) shows the stack combined results for ten randomly initialized networks. The solid line shows the ROC curve for the predictions' mean, and the dashed line shows the same for the median of the predictions. The solid red line shows the ROC curve for the parallel combined ensemble network. The black curve represents a random choice. \label{fig:stack_comp}}
	\end{figure}
	
	In order to estimate the inherent uncertainty on each model, the test data has been divided into randomly selected 50,000 non-overlapping partitions.  \autoref{fig:roc} shows the ROC curve for each model. RNN and CNN are represented with blue and green curves alongside the inherent uncertainty for one standard deviation. The orange curve shows the mean prediction of these two models, which already indicates a higher generalization power than each component network. Finally, the red curve shows the minimalistic ENN configuration. Although the concatenated latent-feature space's training is minimal, it still reveals improvement beyond the mean prediction. The inner plot of \autoref{fig:roc} zooms into the slice of tagging efficiency within $[0.4,0.7]$ to emphasize this improvement. \autoref{fig:roc} also shows the area under the curve (AUC) value for each curve where the improvement in mean prediction and ENN is also visible.  
	\begin{figure}[!h]
		\includegraphics[scale=0.45]{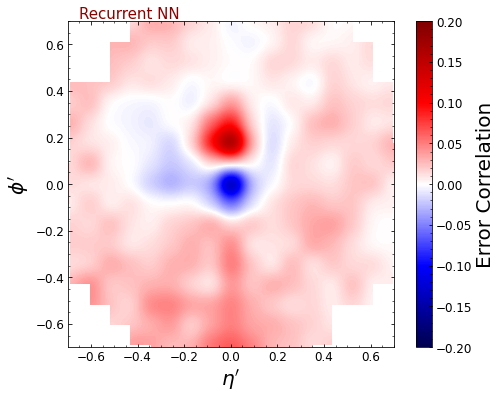}
		\includegraphics[scale=0.45]{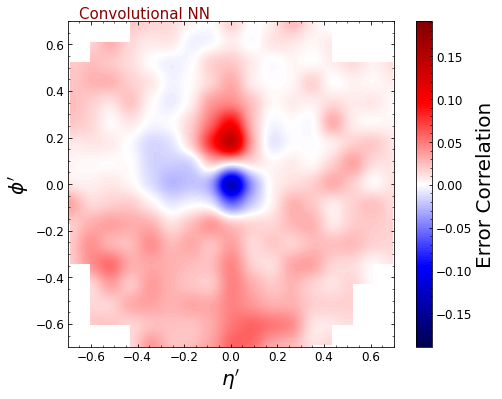}
		\includegraphics[scale=0.45]{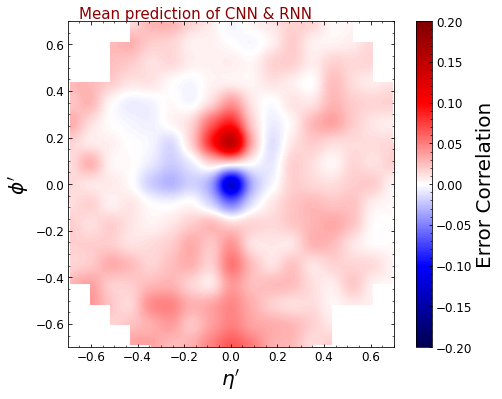}
		\includegraphics[scale=0.45]{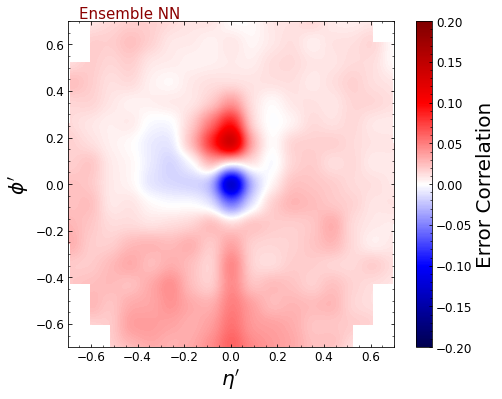}
		\caption{\it Squared error correlation mapped on 50,000 randomly selected test images for RNN (upper left), CNN (upper right), mean (lower left) and ENN (lower right).}\label{fig:errcorr}
	\end{figure}

	As mentioned before, for the ENN to show a significant performance improvement over all pooled networks, it is important for the component networks to show mutually a comparable performance. As seen from \autoref{fig:roc}, both the ENN and the mean prediction is dominated by the CNN above a tagging efficiency of 0.8 and dominated by the RNN below a tagging efficiency of 0.15. This shows that the performance of the network is solely dependent on the correlation between the component networks. In the region $\varepsilon_{S}\in[0.8,1]$ both RNN and CNN captures the 3-prong substructure presented in \autoref{fig:preprocess_image} and \ref{fig:kTsequence} where in the region $\varepsilon_{S}\in[0,0.3]$ a dipole type substructure is captured. Hence, both networks are highly correlated, which reflects in ENN's prediction as well.
	As seen from the interval $[0.4,0.7]$ of the ROC curve, the ENN-improvement is maximized when the component accuracies are similar. 
	
	In ref.~\cite{Kasieczka:2019dbj}, it has been reported that, depending on the architecture, it is possible to improve the prediction quality up to 15\% by using the stack combination method. To assess the parallel combining method's performance with respect to the stack combining method, we retrained RNN and CNN architectures ten times by reinitializing the networks' weights for each training. Then the mean(median)-of-ensemble calculated by combing the predictions of each training. \autoref{fig:stack_comp} shows the comparison between the parallel combined ensemble method (solid red curve) and stacked combined RNN (blue) and CNN (green) architectures. The mean and median combination has been shown with solid and dashed curves. Although we observe a slight improvement in the performance of the stack combined networks with respect to their component network, this improvement does not match with the ENN architecture that this study proposes. We also do not observe a significant difference between using the mean or median for the stack combined methods.
	
	As discussed in \autoref{sec:enns}, combining hypotheses with non-correlated errors may improve an ensemble's prediction. In order to test this, \autoref{fig:errcorr} shows the correlations of the squared error, $(y - \hat{y})^2$ mapped on the test images where $ y $ is the truth label and $ \hat y $ is the prediction of the corresponding network. \autoref{fig:errcorr} shows RNN (upper left panel), CNN (upper right panel), mean prediction (lower left panel) and ENN (lower right panel). Each correlation has been estimated by using randomly selected 50,000 test images. One can immediately see the shrinking area of the blue negative correlation distribution. Although the correlations between the RNN and the CNN mapping look similar, the mean prediction improves the two hypotheses' non-overlapping portions. The ENN goes beyond the mean prediction's achievement by drastically shrinking the blue region and removing the fluctuations in the red (positively correlated) region, hence increasing the correlations between squared error and the image pixels. As expected, similarly correlated regions changed neither for mean prediction nor for ENNs. Thus, combining all available neural networks would not improve the accuracy if their error is highly correlated. Instead, one can benefit from this methodology by employing networks with comparable accuracies and different error correlation to improve the latent-feature space accuracy.

	\section{Bayesian Deep Learning}\label{sec:bnn}
	For all phenomenological applications it is important to assess the intrinsic uncertainties of a NN model. Two major uncertainties can be modelled within the context of DL~\cite{kendall2017uncertainties, Bollweg:2019skg}. The irreducible noise in the observations called aleatoric uncertainties and the uncertainties intrinsic to the proposed hypothesis called epistemic uncertainties. Given sufficient data, epistemic uncertainties can be explained and reduced. The decomposition of the variance of a binary hypothesis is given as~\cite{KWON2020106816,tagasovska2019singlemodel},
	\begin{eqnarray}
	Var(y)\ =\ \underbrace{\langle \hat y ^2 \rangle - \langle \hat y \rangle^2}_{\rm epistemic}\ +\ \underbrace{\langle\ \hat{y}\ (1-\hat{y})\ \rangle}_{\rm aleatoric}\ , \label{eq:unc}
	\end{eqnarray}
	where $\hat y$ represents the network's predictive distribution, the first term represents the epistemic uncertainties while the second term is the aleatoric uncertainty. In addition to the uncertainties, the entropy of the network's prediction, also, gives strong indications about the underlying uncertainties of the system where higher entropy points to higher uncertainty. The entropy of binary classification is given as~\cite{10.5555/971143},
	\begin{eqnarray}
	\mathcal{S} = -\left( \hat{y}\log_2{(\hat{y})} + (1-\hat{y})\log_2{(1-\hat{y})}) \right)\ , \label{eq:entropy}
	\end{eqnarray}
	where the first term stands for the classification of the class 1 (top signal) and the second term for the classification of class 0 (dijet background). 
	
	In order to analyse the uncertainties underlying our neural network, we used the \textsc{TensorFlow Probability} package version 0.10.0~\cite{DBLP:journals/corr/AbadiBCCDDDGIIK16}. We limited ourselves to prediction uncertainties by only changing each network's output layer to Dense Flipout layer~\cite{DBLP:journals/corr/abs-1803-04386} with sigmoid activation\footnote{It is important to note here that, to get the complete model uncertainties from each layer, one can modify the entire network with Bayesian layers. This will double the number of trainable parameters in each layer. Thus in order to simplify our problem, we are only concentrating on prediction uncertainties.}. The kernel divergence function has been chosen to be mean Kullbeck-Leiber divergence. We employed the same network architectures presented in \autoref{sec:nn_achi}. As before, all networks are trained for 500 epochs with \texttt{Adam} optimizer. The initial learning rate has been given as $10^{-4}$ and reduced to its half in every 20 epochs if validation loss has not been improved. The final prediction has been reported using randomly chosen 100,000 test samples where each network output has been sampled 100 times.
	
	Although the notion of ``mean prediction" is ambiguous in the Bayesian context, in order to have a baseline, we defined the mean prediction of CNN and RNN networks as the mean of each 100 samples. This serves as the linear combination ensemble baseline which has not been trained on any latent-feature space beyond its component networks. To reveal our ensembling technique's full effect, we used an ensemble learner with one dense layer including 96 nodes, as before, and another ensemble learner with an additional dense layer with 96 nodes\footnote{It is important to note that we did not observe a significant improvement over ROC AUC by adding an extra dense layer. Thus further optimization beyond adding an extra layer required to improve the accuracy of an ensemble learner. Since this is beyond our scope, we limit ourselves to simplistic architecture.}. 
	\begin{figure}[!h]
		\includegraphics[scale=0.4]{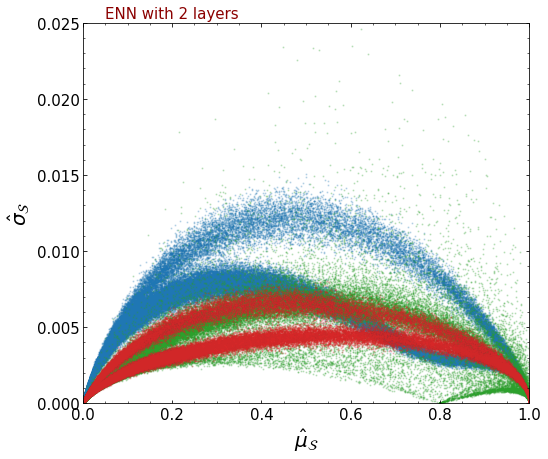}
		\includegraphics[scale=0.4]{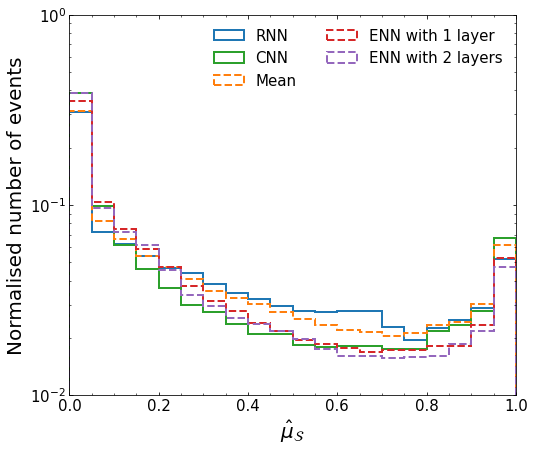}
		\caption{\it Mean entropy distribution with respect to the standard deviation of the entropy for RNN (blue), CNN (green) and ENN (red) where ensemble having two dense layers (left panel). Mean entropy distribution with respect to percentage of binned events (right panel).}\label{fig:entropy}
	\end{figure}
	\begin{table}[!h]
		\centering
		\renewcommand{\arraystretch}{1.5}
		\begin{tabular}{l|ccccc}
			& RNN & CNN & Mean & ENN (1 layer) & ENN (2 layers) \\\hline\hline
			$ \hat{\mu}_S<0.5 $ & $ 71.92\% $ & $ 75.22\% $ & $ 72.61\% $ & $ 78.05\% $ & $ 79.55\% $\\
		\end{tabular}
		\caption{\it Percentage of events for each network structure, i.e. RNN, CNN, ENN and Mean, with mean entropy below 0.5.\label{tab:entropy}}
	\end{table}

	The left panel of \autoref{fig:entropy} shows the mean entropy, $\hat{\mu}_{\mathcal{S}}$, distribution with respect to the standard deviation in entropy, $\hat{\sigma}_{\mathcal{S}}$, where RNN, CNN and two-layer ENN has been represented with blue, green and red points. In order to simplify the plot, the mean prediction and the one-layer ENN model is not shown. One can immediately conclude that the ensemble learner has a considerable limitation on the standard deviation of the entropy where CNN reaches beyond 0.025, RNN to 0.015 but ENN limits the standard deviation below 0.0075. The right panel of \autoref{fig:entropy} shows the percentage of events per mean entropy. As before, the RNN and CNN architectures are represented by blue and green solid curves. The separation between two curves increases between the entropy values $0.2-0.8$ where RNN has been observed to have more events with mid-range entropy values than CNN, but the last bin reveals that CNN has more events with maximum entropy. The dashed orange curve represents the mean of the two predictions where only slight improvement has been observed beyond the RNN. Furthermore, for the two ensemble learners, represented by dashed red and purple curves, one can immediately see the reduction in the number of events for the mid-range entropy values. One can also see that when sufficient complexity is provided, an ensemble learner further improves the hypothesis's entropy, i.e. reduces its values for both $\hat{\mu}_\mathcal{S}$ and $\hat{\sigma}_\mathcal{S}$. This is also summarized in \autoref{tab:entropy}, where more than 78\% of the events for both ensemble learners reach a mean entropy $\hat{\mu}_\mathcal{S}$ of less than 0.5, while RNN, CNN, and mean prediction remain below 75.3\%.
	\begin{figure}[!h]
		\includegraphics[scale=0.4]{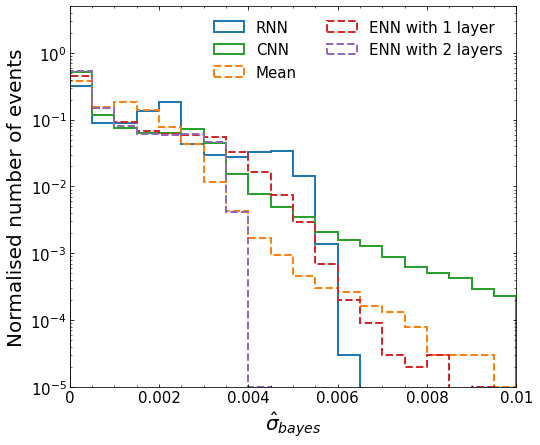}
		\includegraphics[scale=0.4]{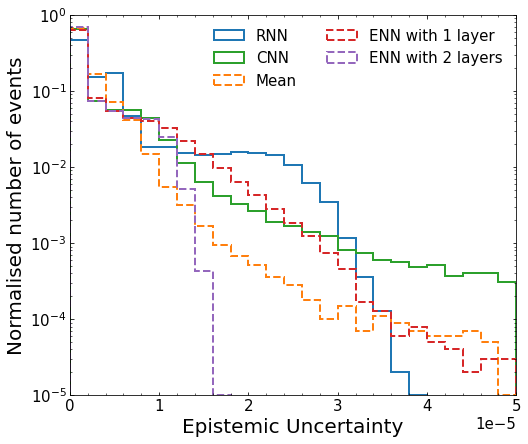}
		\caption{\it Left panel shows the normalised number of events per standard deviation in prediction. Right panel shows the same for epistemic uncertainty. In each histogram RNN, CNN, mean, one-layer ENN and two-layer ENN has been represented with blue, green, orange, red and purple curves.}\label{fig:histos}
	\end{figure}

	We also analyzed the standard deviation in the hypothesis prediction, which is crucial to maintain small in order to achieve consistent predictions. The left panel of \autoref{fig:histos} shows the fraction of events per standard deviation in prediction where the same colour scheme applied as before. Given a sufficiently complexity problem, the ENN is observed to reduce $\hat\sigma_{bayes}$ significantly, compared to each component network and the mean combination of those networks respectively. While the mean prediction reaching up to $\hat\sigma_{bayes}\sim0.01$, the ENN limits the standard deviation below 0.004, which is similar to the standard deviation mean entropy.  On the right panel of the \autoref{fig:histos}, we show the epistemic uncertainty as given in the first term of \autoref{eq:unc} using the same colour labelling. Again, we find a significant reduction of the uncertainties with ensemble learners. These results show that learning over various symmetries leads to a more accurate representation of the given problem without requiring more data.  

\begin{figure}[!h]
		\centering
	\includegraphics[scale=0.4]{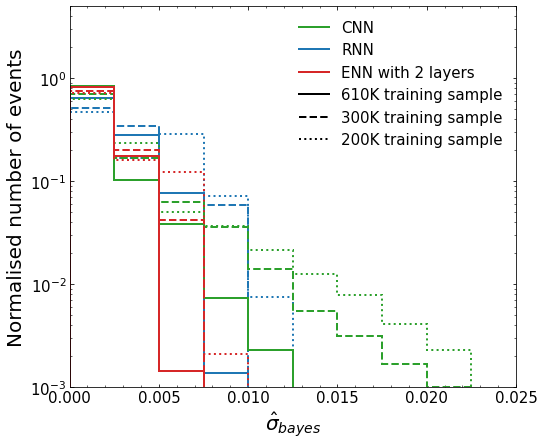}
	\caption{\it The histogram shows the effect of using different training sample sizes on the standard deviation of the prediction. Red, green and blue curves show ENN with two layers, CNN and RNN architectures and solid, dashed and dotted curves shows the training sample sizes of 610,000, 300,000 and 200,000, respectively.}\label{fig:sample_size}
\end{figure}

	An optimization problem requires a sufficient amount of training examples in order to be able to generalize the given hypothesis successfully. As shown in ref.~\cite{Bollweg:2019skg}, lack of variety in training examples will cause uncertainty and the standard deviation of the prediction to increase. \autoref{fig:sample_size} shows the change in the standard deviation in the prediction of CNN (green), RNN (blue) and two-layer ENN (red) architectures. As before, each network output has been sampled 100 times for 100,000 test samples. The solid, dashed, and dotted lines show each network's prediction trained with 610,000, 300,000 and 200,000 randomly chosen training samples. It has been observed that while $\hat\sigma_{bayes}$ gets significantly larger in RNN and CNN architectures, ENN is less susceptible to the lack of training examples. This shows that the ability to access different symmetries within a given data provides the necessary tools for ENN architecture to generalize the hypothesis better with fewer training examples.

	Thus we observed that employing different domains of data that are specialised for specific properties, and optimising a neural network with combined properties of these component learners drastically reduces the system's uncertainties. Such an ensemble network has been shown to learn the system's correlations much more accurately compared to its individual component networks.

	\section{Conclusion}\label{sec:conclusion}
	We presented Ensemble Neural Networks for the reconstruction and classification of collider events and applied them to the discrimination of boosted hadronically decaying top quarks from QCD jets. 
	An ENN can improve the accuracy of the network beyond the individual contributions of its component networks by reducing the variance of the prediction given that the errors of component networks are not highly correlated. In this study, we showed that such techniques can be used in the event reconstruction of collider events in order to overcome the representation problem of neural networks and to improve the prediction accuracy and uncertainties.
	
	Special-purpose networks, such as CNNs or RNNs have been repeatedly shown to be highly accurate for the classification of LHC events. These networks are specialised to learn and optimise their models with respect to the correlations of the given data. In the case of the classification of fat jets, these correlations can be represented through calorimeter images where a network learns the spatial distribution of a jet's constituents. On the other hand, clustering algorithms produce a sequential tree structures which can be employed to learn distinct kinematic features of top decays and QCD backgrounds. An ensemble learner is a paradigm that allows the combination of these properties in one algorithm. Instead of optimising the network separately with respect to the distinct symmetries of images or cluster sequences, it allows optimisation through combined latent-feature space. We showed that combining convolutional and recurrent neural networks and training the network further over their latent-feature space leads to higher accuracy for the classification task. Further, we found that such technique explicitly reduces the variations in error correlations of the component networks hence improving the domains where the component networks are not highly correlated. 

	Although ENN comes with a great advantage, it is crucial to emphasize the trade-off of building such an architecture. ENN is only valid if its component networks can capture different correlations in the data. As shown in \autoref{sec:nn_achi}, ENN can not improve the regions where the errors of the components are highly correlated. Hence, in such cases, it would be equally beneficial to focus on improving the performance of individual state-of-the-art NN architectures. It is also important to note that this does not render the stack combining method invalid. For complex loss hypersurfaces, it is quite challenging for a learning algorithm to find the global minimum. If there is no other architecture available that can exploit different features of the given data, then the stack combining method will achieve a much closer approximation by sampling different regions of the hypothesis-space.
	
	A detailed understanding of the inner workings of Deep Learning techniques is often missing. To develop confidence in their applicability in measurements and searches for new physics, it is of vital importance to understand and, if possible, reduce the uncertainties of the networks. Bayesian techniques are designed to quantify such uncertainties. We found that ENNs can drastically reduce the uncertainty in the prediction of the network, without increasing the amount of training data.  We also showed that such methods reduce the entropy of the system as well as the epistemic uncertainties and it reduces the network's susceptible to small-sized training samples. ENNs can thus provide much more accurate predictions than their component networks. The methodology employed in this study can be applied to a broad scope of application in HEP phenomenology. Instead of expanding the data domain, learning through combined underlying correlations of the problem has been shown to be very effective.
	
	While ensemble learners can reduce the variance of the hypothesis, we did not observe any improvement in the data's bias or aleatoric uncertainties. Although reducing the network's epistemic uncertainties and variance is a crucial step, aleatoric uncertainties are observed to be larger than the epistemic uncertainties. As it has been shown that Genetic-Algorithm-based Selective Ensembles can reduce the biases as well as the variance of the system~\cite{ZHOU2002239}, it is an obvious next step to employ such techniques to reduce biases as well as the variance of the network.

	\bibliography{bibliography}
\end{document}